\documentclass[twocolumn,american]{article}
\usepackage{pslatex}
\usepackage[T1]{fontenc}
\usepackage[latin1]{inputenc}
\usepackage{geometry}
\usepackage{amsmath}
\usepackage{graphicx}
\usepackage{amssymb}

\makeatletter

\providecommand{\tabularnewline}{\\}

\date{}

\AtBeginDocument{
  
}

\usepackage{babel}
\makeatother
\begin{document}

\title{The Thorium Molten Salt Reactor :\\
Moving On from the MSBR}

\author{{\normalsize L. Mathieu, D. Heuer, R. Brissot, C. Le Brun, E. Liatard,
J.-M. Loiseaux,}\\
{\normalsize O. Méplan, E. Merle-Lucotte, A. Nuttin and J. Wilson,}\\
{\normalsize Laboratoire de Physique Subatomique et de Cosmologie
,}\\
{\normalsize 53, avenue des Martyrs, F-38026 Grenoble Cedex, France}\\
{\normalsize }\\
{\normalsize C. Garzenne, D. Lecarpentier, E.Walle,}\\
{\normalsize EDF-R\&D, Département SINETICS,}\\
{\normalsize 1 av du Général De Gaulle, 92140 Clamart, France}}

\maketitle
\begin{abstract}
A re-evaluation of the Molten Salt Breeder Reactor concept has revealed
problems related to its safety and to the complexity of the reprocessing
considered. A reflection is carried out anew in view of finding innovative
solutions leading to the Thorium Molten Salt Reactor concept. Several
main constraints are established and serve as guides to parametric
evaluations. These then give an understanding of the influence of
important core parameters on the reactor's operation. The aim of this
paper is to discuss this vast research domain and to single out the
Molten Salt Reactor configurations that deserve further evaluation.
\end{abstract}

\section*{Introduction}

In order to reduce $\textrm{CO}_{2}$ emissions in the coming decades,
and, as a result, to mitigate global warming, it appears necessary
to stabilize or, better, to reduce the use of fossil fuels. Resorting
to a sustainable version of nuclear power may help replace classical
energy production partially and thus satisfy an increasing world energy
demand while conserving the climate and natural resources. The Generation
IV International Forum for the development of new nuclear energy systems
\cite{Gen-IV} has established a set of goals as research directions
for nuclear systems: enhanced safety and reliability, reduced waste
generation, effective use of uranium or thorium ores, resistance to
proliferation, improved economic competitiveness. Molten Salt Reactors
(MSR) are one of the systems retained by Generation IV. MSRs are based
on a liquid fuel, so that their technology is fundamentally different
from the solid fuel technologies currently in use. Some of the advantages
specific to MSRs (in terms of safety/reliability, for example) originate
directly from this characteristic \cite{RSF}. Furthermore, this type
of reactor is particularly well adapted to the thorium fuel cycle
(Th-$^{233}\textrm{U}$) which has the advantage of producing less
minor actinides than the uranium-plutonium fuel cycle ($^{238}\textrm{U}$-$^{239}\textrm{Pu}$)
\cite{Lecarpentier,Alex}. Moreover, while breeding or regeneration
in the U-Pu cycle can be obtained only with a fast neutron spectrum,
in the Th-$^{233}\textrm{U}$ fuel cycle it can, in principle, be
obtained with a more or less moderated neutron spectrum. In a thermal
neutron spectrum, poisoning due to the Fission Products (FP) being
worse than in a fast neutron spectrum, the rate at which fuel reprocessing
is performed can become a major issue. Because, in an MSR, the fuel
is liquid, continuous extraction of the FPs is a possibility. Although
MSRs can be operated as incinerators they will be discussed in this
paper only as electricity producing critical systems.

In 1964, the Molten Salt Reactor Experiment (MSRE) was initiated at
the Oak Ridge National Laboratory (ORNL). Generating 8 MWth of power,
the reactor was operated without problems and with different fuels
($^{235}\textrm{U}$ then $^{233}\textrm{U}$) over several years.
The expertise gained during this experiment led, in the 1970s, to
the elaboration of a power reactor project, the Molten Salt Breeder
Reactor (MSBR \cite{MSBR}). The studies demonstrated that fuel regeneration
is possible with the thorium fuel cycle in an epithermal spectrum,
provided very efficient and, as a consequence, constraining, on-line
chemical reprocessing of the salt is achieved. Over the past few years,
the MSBR has been reassessed in the light of new calculating methods
\cite{Lecarpentier,Alex} so as to elaborate a new reactor concept
that we call the Thorium Molten Salt Reactor (TMSR).

The MSBR suffered from several major drawbacks and was discontinued.
The goal being, at the time, to obtain as high a breeding ratio as
possible, the on-line chemical reprocessing unit considered had to
process the entire salt volume within 10 days and this was very complex
\cite{Walle}. Because of this complexity, the project is often considered
unfeasible. In addition, recent calculations have shown that the global
feedback coefficient for this system is slightly positive. This contradicts
the results that had been presented. The difference is probably due
to the fact that, at the time, the compositions were handled in a
homogeneous way while they are handled heterogeneously today. This
critical issue makes the MSBR a potentially unstable system in some
situations.

The aim of this paper is to present solutions to these problems. In
our search for better reactor configurations, we have identified several
constraints that are discussed in the first part of this work. We
then discuss the impact of the various reactor parameters on these
constraints, i.e. chemical reprocessing, channel size, fuel volume
and the proportion of heavy nuclei (HN) in the salt. A synthesis of
these studies is set forth in the last section.

This work is based on the coupling of a neutron transport code (MCNP
\cite{MCNP}) with a materials evolution code. The former calculates
the neutron flux and the reaction rates in all the cells while the
latter solves the Bateman equations for the evolution of the materials
composition in the cells. These calculations take into account the
input parameters (power released, criticality level, chemistry,...),
by adjusting the neutron flux or the materials composition of the
core on a regular basis. Our calculations are based on a precise description
of the geometry and consider several hundreds of nuclei with their
interactions and radioactive decay; they allow fine interpretation
of the results. All the data discussed in this paper result from the
evolution of the reactor over 100 years.

\section{Constraints}

We identify five major constraints in this study: safety, chemical
reprocessing feasibility, fuel regeneration capability, materials
life span, and initial inventory. Other constraints could be considered,
such as waste minimization, thermal-hydraulics, or proliferation resistance
but we concentrate essentially on the above five major constraints.
We seek to understand the impact of the reactor's defining parameters
on these constraints. In so doing, we can single out the best reactor
configurations according to the weight assigned to each of the constraints.

\subsection{Safety}

In the work we present here this constraint concerns essentially the
evolution of the feedback coefficients that should be negative. The
more kinetic aspects of the reactor's safety properties are not considered
here. Additionally, the ways in which we change the concept do not
modify the other MSR safety properties, such as fuel dumping and the
fact that MSRs are free of high pressure areas.

The feedback coefficients, $\frac{dk}{dT}$, are a measure of the
variation of the multiplication factor ($dk$) with the temperature
of the core or of a portion of the core ($dT$). The global feedback
coefficient can be broken up into several strongly uncorrelated partial
coefficients, each of which characterizes the variation of a specific
parameter: the effects due to the expansion of the salt%
\footnote{~The heating of the salt induces a widening of the resonances due
to the Doppler effect and a change in the neutron spectrum moderation
due to the salt. Both of these effects are considered together, under
the term {}``Doppler''.%
}, and the purely thermal effects of the salt and of the graphite.
This reads:

\[
\left.\frac{dk}{dT}\right)_{total}=\left.\frac{dk}{dT}\right)_{density}+\left.\frac{dk}{dT}\right)_{Doppler}+\left.\frac{dk}{dT}\right)_{graphite}\]

In order for the reactor to be intrinsically safe, a temperature increase
must not induce an increase of the reactivity and, as a consequence,
of the power released by fissions. For this reason, the $\left.\frac{dk}{dT}\right)_{total}$
coefficient must be negative. The thermal kinetics of the graphite,
which is heated by gamma radiation and cooled by the salt, is much
slower than that of the salt. Making allowance for this delay between
the heating of the salt and the heating of the graphite, the coefficient
for the salt alone, that is the sum $\left.\frac{dk}{dT}\right)_{density}+\left.\frac{dk}{dT}\right)_{Doppler}$
must also be negative. The degree of safety can be further increased
if the density coefficient is made negative. This implies that any
local loss of density, e.g. because of a bubble, decreases the reactivity
of the system.

The uncertainties on these values are related to statistical errors
that are well identified and can be reduced, but also to systematic
errors that are not quantified and are related, for example, to uncertainties
in cross section evaluations. For this reason, the feedback coefficients
must be sufficiently negative to ensure unambiguous stability.

\subsection{Feasibility of the Chemical Reprocessing}

The term feasibility reflects the complexity associated to the chemical
reprocessing. Indeed, some of the separation processes are considered
too difficult to be implemented. This can have several causes: the
processes considered are not well understood or mastered, the flow
of materials to be processed is too large, the reprocessing implies
direct coupling to the reactor core,...

The objective is to devise the simplest possible system that is compatible
with the other constraints. In particular, it will be important to
avoid excessive deterioration of the system's fuel regeneration capability.

\subsection{Fuel Regeneration Capability}

The breeding ratio expresses the balance between the creation of $^{233}\textrm{U}$
through neutron capture on $^{232}\textrm{Th}$ and the destruction
of $^{233}\textrm{U}$ through fission or neutron capture. The breeding
ratio in a critical reactor can thus be written:

\[
BR=\frac{r_{c,^{232}Th}-r_{c,^{233}Pa}}{r_{f,^{233}U}+r_{c,^{233}U}}\]

With $r_{c}$ and $r_{f}$ respectively the capture rate and the fission
rate of the different isotopes.

A breeding ratio less than 1 implies that $^{233}\textrm{U}$ is consumed
so that fissile matter must be fed into the core on a regular basis.
This inevitably increases both the volume and the frequency of transfer
of these dangerous materials. Similarly, a breeding ratio larger than
1 implies that the excess $^{233}\textrm{U}$ produced be placed in
storage and/or transported. Because, in all cases, the initial fissile
matter inventory has to be produced by other means (e.g. in pressurized
water reactors or fast neutron reactors) the highest possible breeding
ratio does not necessarily have to be sought.

In order to satisfy the regeneration constraint, we try to achieve
a breeding ratio at least equal to 1, knowing that any excess neutrons
can always be put to use (improved safety, transmutation capabilities,
... ).

\subsection{Materials Life Span}

This concerns in particular how the graphite reacts to irradiation
exposure. Beyond a certain degree of damage, it becomes the seat of
swelling. Graphite's life span is determined by the time it takes
to reach a fluence limit, that we will set to $2.10^{22}$~n/$\textrm{cm}^{2}$
at a temperature of 630~°C \cite{tenue_graphite}. In our calculation,
we consider only the neutrons whose energy is larger than 50~keV,
i.e. those that create real damage in the graphite. 

The goal, with this constraint, is to obtain a life span that is not
too short so as to avoid replacing the core graphite too frequently.

\subsection{Initial Inventory}

The inventory, here, is the amount of $^{233}\textrm{U}$ needed to
start a 1~GWe power reactor. The smaller the inventory, the faster
the deployment of a fleet of such reactors can be achieved \cite{Elsa-LPSC,Elsa-Physor}.

Without excluding configurations with a large inventory, its minimization
will be sought.\\

These constraints are not all equivalent; a weighting factor can be
assigned to each of them. This factor depends on the technologies
available and the goals that guide reactor choices. As the performance
of a system depends on how the constraints are weighted and on how
difficult it is to satisfy them, it is not possible to specify the
{}``best'' solution. The only possibility is to identify a number
of interesting trends. This yields a better understanding of the system
and can lead to the definition of a power reactor (stringent constraints)
or of a demonstration unit (less stringent constraints).

\section{Impact of the Parameters on the Constraints}

In this section, we examine how various reactor parameters impact
the five constraints discussed above. In order to be able to compare
the systems studied, we found it useful to define a standard system
from which the different studies could stem.

Our standard system is a 1~GWe graphite moderated reactor. Its operating
temperature is 630~°C and its thermodynamic efficiency is 40~\%.
The graphite matrix comprises a lattice of hexagonal elements with
15~cm sides. The total diameter of the matrix is 3.20~m. Its height
is also 3.20~m. The density of this nuclear grade graphite is set
to 1.86. The salt runs through the middle of each of the elements,
in a channel whose radius is 8.5~cm. One third of the 20~$\textrm{m}^{3}$
of fuel salt circulates in external circuits and, as a consequence,
outside of the neutron flux. A thorium and graphite radial blanket
surrounds the core so as to improve the system's regeneration capability.
The properties of the blanket are such that it stops approximately
80~\% of the neutrons, thus protecting external structures from irradiation
while improving regeneration. We assume that the $^{233}\textrm{U}$
produced in the blanket is extracted within a 6 month period.

The salt used is a binary salt, LiF - $\textrm{(HN)F}_{4}$, whose
$\textrm{(HN)F}_{4}$ proportion is set at 22~\% (eutectic point),
corresponding to a melting temperature of 565~°C. The salt density
at 630~°C is set at 4.3 with a dilatation coefficient of $10^{-3}$/°C
\cite{Walle2}. We assume that helium bubbling in the salt circuit
is able to extract the gaseous fission products and the noble metals
within 30 seconds. The standard reprocessing we consider is the delayed
reprocessing of the total salt volume over a 6 month period with external
storage of the Pa and complete extraction of the FPs and the TRansUranians
(TRU) (Figure \ref{cap:Reprocessing-scheme}).

\subsection{Influence of the Reprocessing}

\subsubsection{How Slow Delayed Reprocessing Works}

As previously stated, the MSBR reprocessing is considered too complex
to be feasible in the next few decades. The effectiveness of this
reprocessing rested mainly on the extraction and storage of the protactinium
away from the neutron flux so as to avoid, insofar as possible, the
production of $^{234}\textrm{U}$ by neutron capture. The half-life
of $^{233}\textrm{Pa}$ is 27 days and its extraction has to be markedly
faster if it is to be efficient. That is why the reprocessing of the
total core volume in 10 days was contemplated.

\begin{figure*}[t]
\begin{center}\includegraphics[%
  bb=14bp 14bp 836bp 476bp,
  clip,
  scale=0.5]{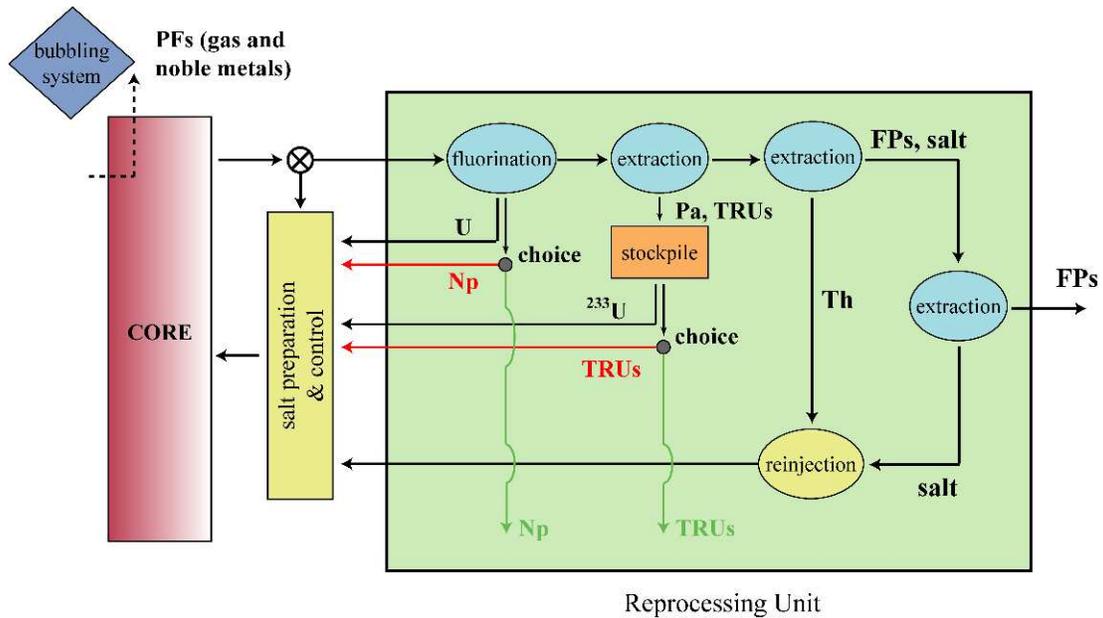}\end{center}

\caption{Slow reprocessing overview\label{cap:Reprocessing-scheme}}
\end{figure*}

The difficult part of the reprocessing is Fission Product extraction
in the presence of thorium. The idea, with slow reprocessing, is to
first extract the thorium, so as to avoid being handicapped by its
presence in the FP extraction process. This method could not be applied
in the MSBR because of the large thorium flow involved, reaching several
tons per day while it is only a few hundreds of kilograms per day
in the case of a six month reprocessing time.

In addition, with slow reprocessing, the nuclear core can be disconnected
from the processing unit, small amounts of the salt being processed
individually, instead of resorting to continuous on-line reprocessing,
as in the MSBR. This is a source of simplification, it allows easier
control of the procedure while making the core less sensitive to possible
problems in the reprocessing unit.

Figure \ref{cap:Reprocessing-scheme} gives a general view of what
slow reprocessing could entail. Some of the stages shown in this general
schematic, such as protactinium storage, can be eliminated while maintaining
the primary assets of the reprocessing. Likewise, the neptunium extracted
in the course of the first fluorination, and the other TRansUranians
can be either reinjected in the core or managed separately. The advantage,
in the first option, is that an {}``incinerating'' configuration
is obtained, insofar as all the TRUs are kept in the core. With the
second option, the production of americium, curium, and other heavier
elements is significantly reduced.

The time allocated to cleaning the salt and reinjecting it can be
extended considerably. Indeed, if the time needed to reprocess the
core volume is equal to the time before reinjecting the salt, there
is as much salt outside the core as inside it. Thus, up to 6 months
can separate the extraction of the fuel salt and its reinjection in
the core, after removal of the FPs. The fissile matter inventory is
not increased, however, thanks to the possibility of extracting the
uranium during a preliminary fluorination stage. In the case of slow
reprocessing, we assume very good extraction efficiencies (they are
set to 1 in the calculations) because plenty of time is available.

It is too early to say that such a reprocessing scheme solves the
feasibility issues; it is, however, possible to assert that the simplification
of the system improves its feasibility. The impact of this reprocessing
on the other constraints, in particular those of regeneration and
the feedback coefficients has to be assessed.

\subsubsection{Impact of the Reprocessing Time}

In Table \ref{cap:Breeding-ratio-fonction-reprocessing}, the breeding
ratios obtained at equilibrium are given for various reprocessing
options as applied to the reactor configuration described previously.
The best breeding ratio is obtained with the MSBR reprocessing and
the worst with no reprocessing other than helium bubbling in the core,
and $^{233}\textrm{U}$ recovery in the blanket.

In the table, the MSBR reprocessing is labeled {}``fast (10 days)''
because of the rate at which the protactinium is to be extracted.
However, the extraction of the FPs is partial, making the real reprocessing
rate longer (equivalent to 50 days for the FPs that capture the most).
The option labeled {}``bubbling only'' is set apart because it is
dramatically different from the other configurations, making any comparison
with them tricky~(no equilibrium state).

\begin{table}
\vspace{5mm}
\begin{center}\begin{tabular}{|c|c|c|}
\hline 
Reprocessing time&
Breeding ratio&
dk/dT (pcm/°C)\tabularnewline
\hline
\hline 
Fast (10 days)&
1.062&
-2.25\tabularnewline
\hline 
Slow (3 months)&
1.024&
-2.37\tabularnewline
\hline 
Slow (6 months)&
1.000&
-2.36\tabularnewline
\hline 
Slow (1 year)&
0.986&
-2.39\tabularnewline
\hline 
Slow (2 years)&
0.961&
-2.50\tabularnewline
\hline
\hline 
Bubbling only&
0.562&
-2.2\tabularnewline
\hline
\end{tabular}\end{center}

\caption{Breeding ratio and feedback coefficient for several reprocessing
options. The statistical error on the feedback coefficient is less
than 0.05 pcm/°C.\label{cap:Breeding-ratio-fonction-reprocessing}}
\end{table}

Varying the reprocessing time from 3 months to 2 years induces about
a 0.06 loss in the breeding ratio. For these four configurations,
the proportion of protactinium stored outside of the neutron flux
is, respectively, 30~\%, 20~\%, 10~\% and 5~\%. However, the change
in the breeding ratio is due mainly to the change in the capture rate
of the FPs and, to a lesser degree, of the TRUs. On the contrary,
with fast reprocessing, 80~\% of the protactinium is stored outside
of the neutron flux and that is the direct cause of the system's good
breeding ratio, way before the FPs and the TRUs. Thus, unless it is
extracted rapidly, the Pa's incidence on regeneration is minor.

We now know the leeway afforded by the reprocessing, since a doubling
of the reprocessing time induces a breeding ratio loss of about 0.02.
To be precise, we should add that the degradation of the breeding
ratio is three times smaller in configurations with a fast neutron
spectrum, where the proportion of graphite in the core is reduced.

It is important to note that the reprocessing option chosen has a
moderate impact on the feedback coefficients, as shown in the table.
This means that reprocessing time and safety can, in a first approximation,
be considered to be independent.

\subsubsection{Destination of the TRansUranians}

As previously stated, the TRUs can either be fed back into the core
or they can be managed separately (incinerated in sub-critical reactors,
incinerated in fast neutron reactors, or placed in storage). The choice
has an impact on the regeneration capabilities, as shown in Table
\ref{cap:Breeding-ratio-fonction-TRU}. Indeed, even if some of the
TRUs fission, they impair the neutron balance because of their high
capture rates. In the auto-incinerating configuration, the most capturing
TRUs reach equilibrium within about 30 years and contribute to the
deterioration of the neutron balance.

\begin{table}[h]
\vspace{5mm}
\begin{center}\begin{tabular}{|c|c|c|}
\hline 
Reprocessing&
Breeding ratio&
dk/dT (pcm/°C)\tabularnewline
\hline
\hline 
TRUs extracted&
1.000&
-2.36\tabularnewline
\hline 
TRUs reinjected&
0.987&
-3.12\tabularnewline
\hline
\end{tabular}\end{center}

\caption{Breeding ratio and feedback coefficient according to TRU management.
The statistical error on the feedback coefficients is less than 0.05~pcm/°C.\label{cap:Breeding-ratio-fonction-TRU}}
\end{table}

In the same table, the influence of the TRUs on the feedback coefficients
is also shown. This coefficient is slightly improved if the TRUs are
kept in the core. This is because the TRUs harden the spectrum, as
will be discussed further in Section \ref{sub:Influence-de-la-taille_des_canaux}.
Note that, in a fast neutron spectrum configuration, the impact of
TRU reinjection on both the breeding ratio and the feedback coefficients
is reduced. 

TRU extraction, however, is advantageous in terms of waste production.
When they are submitted to a neutron flux, TRUs form, progressively,
significant amounts of very heavy elements such as curium. The ratio
of capture to fission cross sections is not favorable to incineration
in this type of reactor because of its epithermal neutron spectrum.
If these elements are removed from the reactor core, larger amounts
of neptunium, formed constantly by captures on $^{236}\textrm{U}$,
are extracted, but the production rate of the other actinides is reduced,
as shown in Table \ref{cap:Table-des-TRU}. The goal, then, is to
obtain TRUs that are more manageable in view of incorporating them
in the fuel of Fast Neutron Reactors. If such an outlet for TRUs is
not available, this option is of no interest.

\begin{table}[h]
\vspace{5mm}
\begin{center}\begin{tabular}{|c|c|c|c|}
\hline 
&
TRUs reinjected&
\multicolumn{2}{c|}{TRUs extracted}\tabularnewline
&
(inventory)&
(inventory)&
(output flow)\tabularnewline
\hline
\hline 
Np&
105 kg&
15 kg&
4.3 kg / TWh\tabularnewline
\hline 
Pu&
265 kg&
2.7 kg&
770 g / TWh\tabularnewline
\hline 
Am&
7.2 kg&
0.5 g&
0.14 g / TWh\tabularnewline
\hline 
Cm&
17.5 kg&
0.1 g&
30 mg / TWh\tabularnewline
\hline
\end{tabular}\end{center}

\caption{TRU production and in core inventory at the end of the time period
covered by this study (100 years) for two TRU destinations; reprocessing
time is 6 months. The output flow calculation is based on 7~TWh per
year energy produced.\label{cap:Table-des-TRU}}
\end{table}

\subsection{Influence of the Size of the Channels\label{sub:Influence-de-la-taille_des_canaux}}

The size of the channels in which the salt circulates is a fundamental
parameter of the reactor. Since the size of the hexagons is kept constant
in all of our studies, the size of the channels determines the moderation
ratio. Changing the radius of the channels modifies the behavior of
the core, placing it anywhere between a very thermalized neutron spectrum
and a relatively fast spectrum.

The two extreme possibilities correspond respectively to a large number
of very small channels and a single big salt channel. In the latter
configuration, there is no graphite in the hexagons and the core consists
in a single channel. In order to allow a comparison of the results
with those of the other configurations, in this case, the hexagons
are treated as salt channels with an equivalent area (channel radius:
13.6~cm.).

For the configurations in which the channel radius is equal to or
larger than 10~cm, it is essential that the graphite of the axial
reflectors be replaced with less moderating materials (e.g. zirconium
carbide). Otherwise, the fissions occur massively in the vicinity
of the reflectors instead of within the core.

\begin{figure}[h]
\vspace{5mm}
\begin{center}\includegraphics[%
  clip,
  scale=0.32]{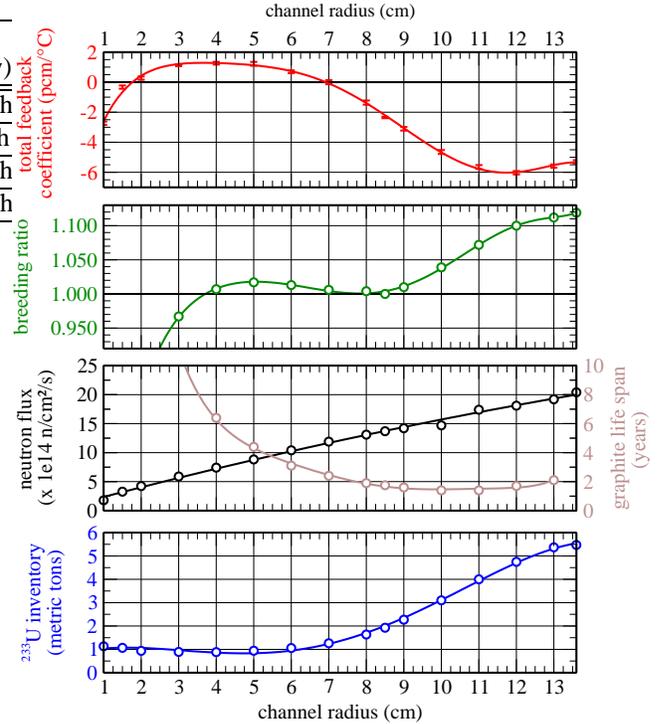}\end{center}

\caption{Influence of the channel radius on four of the five constraints (configuration:
variable radius, 20 $\textrm{m}^{3}$ salt, 630°C, 22\% (HN)$\textrm{F}_{4}$)\label{cap:Influence-of-the-channel-radius}}
\end{figure}

As shown in Figure \ref{cap:Influence-of-the-channel-radius}, the
radius of the channels has a strong impact on most of the constraints%
\footnote{~The moderation ratio can seem to be a more universal parameter but,
like the radius of the channels, it is also influenced by other parameters.
An identical moderation ratio can yield very different results according
to the density of the materials involved or the size of the hexagons.%
}. 

\subsubsection{Safety Constraint}

\begin{figure}[h]
\begin{center}\includegraphics[%
  clip,
  scale=0.3]{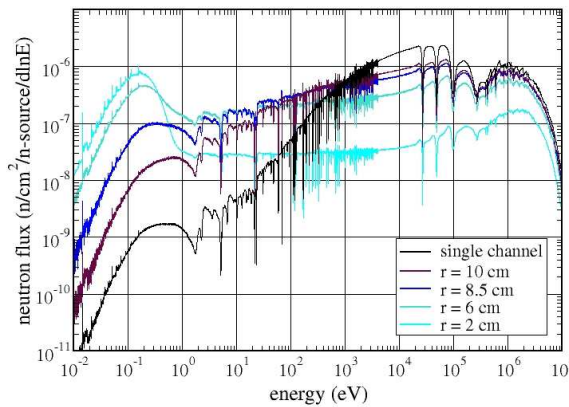}\end{center}

\caption{Neutron spectrum for several channel radii (configuration: variable
radius, 20~$\textrm{m}^{3}$ salt, 630~°C, 22\%~(HN)$\textrm{F}_{4}$)\label{cap:Spectre-de-neutron-fonction-rayon}}
\end{figure}

The study of the feedback coefficients requires a fine analysis of
the neutron spectra involved. These are shown in Figure \ref{cap:Spectre-de-neutron-fonction-rayon}
for different channel sizes up to the single channel configuration.
The cross section resonances of the materials present in the core
have a strong impact on the neutronic behavior of the reactor. The
main resonances are visible: fission of $^{233}\textrm{U}$ at about
2~eV, $^{234}\textrm{U}$ capture at 5~eV, $^{232}\textrm{Th}$
capture near 22~eV, and diffusion on $^{19}\textrm{F}$ near 25,
50 and 100~keV.

\begin{figure}[h]
\vspace{5mm}
\begin{center}\includegraphics[%
  clip,
  scale=0.3]{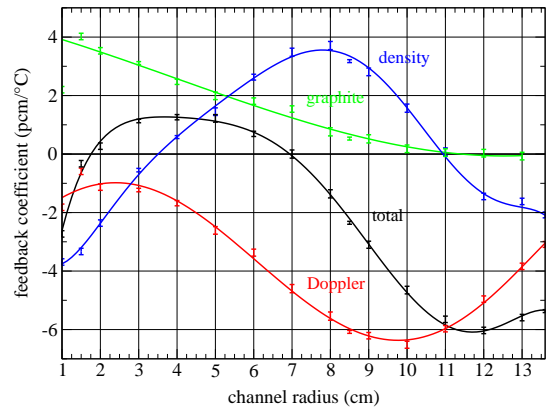}\end{center}

\caption{Total feedback coefficient and feedback coefficient components versus
channel radius (configuration: variable radius, 20~$\textrm{m}^{3}$
salt, 630~°C, 22\%~(HN)$\textrm{F}_{4}$)\label{cap:D=E9composition-du-coefficient-fonction-rayon}}
\end{figure}

As shown in Figure \ref{cap:Influence-of-the-channel-radius}, the
total feedback coefficient becomes rather strongly negative as the
spectrum hardens. This evolution is due to the conjoined variation
of the three sub-coefficients, Doppler, density and graphite, as illustrated
in Figure \ref{cap:D=E9composition-du-coefficient-fonction-rayon}. 

The Doppler coefficient is linked to the $^{233}\textrm{U}$ fission
resonance and the $^{232}\textrm{Th}$ capture resonance (and, to
a lesser degree, to the $^{234}\textrm{U}$ capture resonance). These
two elements have opposite effects on the feedback coefficient: $^{233}\textrm{U}$
worsens it whereas $^{232}\textrm{Th}$ improves it. The thermal agitation
of the salt nuclei induces a widening of these resonances so that
their influence is increased. The value of the Doppler coefficient
depends on how intense the flux is at these resonance values. When
the spectrum hardens, the flux is more intense for high energies and
less so for low energies, the thorium resonances are favored (main
resonance located at 22~eV while that of $^{233}\textrm{U}$ is at
2~eV). The Doppler coefficient then becomes more negative. Beyond
a certain degree of spectrum hardening, the large resonances of both
thorium and uranium lie in a zone where the flux has a low intensity
and their importance is reduced. This explains the worsening of the
Doppler coefficient for large radii.

The density coefficient is related to the expansion of the salt which
pushes a fraction of the fuel outside of the moderated zone. The consequence
is spectrum softening because the proportion of graphite to salt is
larger, thus increasing the fission rate. The effect is small for
small radii where thermalization is already very efficient and where
it is counterbalanced by captures in the graphite. For large radii,
the thermal part of the spectrum contributes practically nothing in
the neutron balance and the effects of neutron escape are felt more
strongly. The density coefficient can become negative when the effects
of captures in the graphite (small radius, large proportion of graphite)
or of neutron escapes (large radius, fast neutron spectrum) dominate
over the effects of thermalization.

The graphite coefficient comes from an energy shift of the thermal
part of the neutron spectrum (around 0.2~eV), due to heating of the
moderator. This shift increases the fission rate because of a small
low energy (0.3~eV) resonance in the fission cross section of $^{233}\textrm{U}$
\cite{Alex}. Its impact on the stability decreases as the amount
of graphite in the core decreases and as the influence of the thermal
portion of the spectrum weakens.

\subsubsection{Regeneration Constraint}

The capacity to regenerate the fuel varies a great deal with the size
of the channels. This can be explained on one hand by the number of
neutrons available and, on the other hand, by the increased neutron
losses in configurations with small channel radii. The number of neutrons
available represents the number of leftover neutrons once both the
chain reaction and the regeneration are ensured. This is defined by:

\[
N_{a}=\nu-2\left(1+\alpha\right)\]
Where $\alpha$ is the mean capture to the mean fission cross section
ratio of $^{233}\textrm{U}$. These available neutrons are distributed
mainly between sterile captures and supplementary captures in thorium
(breeding). $N_{a}$ reaches a minimum at about r = 8.5 cm because
of the variations of $\alpha$ with the neutron spectrum. For small
radii, the strong dip in the breeding ratio is due to neutron losses
in the graphite because there is so much of it.

\subsubsection{Materials Life Span Constraint}

The mean cross sections decrease dramatically with the hardening of
the neutron spectrum. As increasing the inventory does not compensate
for this loss, the neutron flux has to be increased in order to keep
the power constant.

While this phenomenon is linear, this is not true of the core graphite's
life span, as shown in Figure \ref{cap:Influence-of-the-channel-radius}.
A few items have to be stressed:

\begin{itemize}
\item The graphite in the center undergoes a flux that is more intense than
in the periphery. The life span we provide is averaged over the entire
core.
\item The maximum fluence in the graphite decreases when the temperature
increases and the temperature is not uniform. Since the graphite is
heated by gamma radiation and cooled by the salt, the temperature
is higher between channels and lower on the channel surface. The temperature
difference increases as the channels are further apart. This means
that configurations with a smaller channel radius should have a smaller
maximum authorized fluence than those with a larger channel radius.
\end{itemize}
Generally, the flux in the graphite is directly related to the flux
in the salt so that increasing the flux in the salt reduces the graphite's
life span. It is thus considerably shorter with larger channel radii
than with smaller ones.

As can be noticed, the graphite life span curve does not extend all
the way to the single channel configuration since, in that configuration,
there is no graphite inside the core (except that of the blanket).
This configuration then has an asset in that it almost completely
solves the issue of the graphite's life span.

\subsubsection{In Core Inventory Constraint}

For the reactor to be critical, the fissile matter inventory has to
be adjusted when the neutron spectrum hardens. Indeed, the mean $^{233}\textrm{U}$
fission cross section and the mean $^{232}\textrm{Th}$ capture cross
section decrease as the energy of the neutrons increases but the evolution
is not identical for the two isotopes. Two different operating regimes
can be singled out as shown in Figure \ref{cap:Influence-of-the-channel-radius}.

\begin{itemize}
\item For small channel radii, the cross section decreases practically in
the same way for the two isotopes and the inventory required does
not change much.
\item For larger radii, beyond 7~cm, the mean fission cross section of
$^{233}\textrm{U}$ decreases faster than the capture cross section
of $^{232}\textrm{Th}$ so that the inventory has to be increased
significantly.
\end{itemize}

\subsection{Influence of the Salt Volume}

The power per unit volume of salt (specific power) is a determining
parameter in a reactor's behavior. In the reference configuration,
it amounts to about 250~W/$\textrm{cm}^{3}$ for the salt in the
core. This parameter can be modified in two ways: by changing the
fuel volume at fixed power or by changing the total reactor power
at fixed salt volume. These two options yield similar results and
only the first one is discussed in this paper.

\begin{figure}[h]
\vspace{5mm}
\begin{center}\includegraphics[%
  clip,
  scale=0.32]{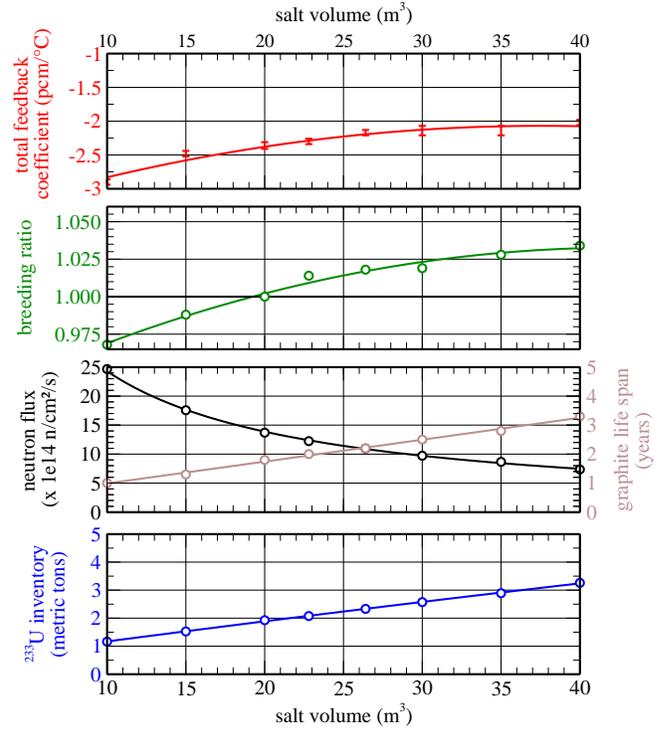}\end{center}

\caption{Influence of the salt volume on four of the five constraints (configuration:
r = 8.5 cm, variable salt volume, 630°C, 22\% (HN)$\textrm{F}_{4}$)\label{cap:Influence-of-salt-volume}}
\end{figure}

Since the flow of Heavy Nuclei is considered to be a key factor for
the feasibility of the chemical reprocessing, the reprocessing time
is adjusted so as to keep this flow constant from one system to the
other. Thus, doubling the salt volume implies that core reprocessing
takes twice as long. The incidence of the salt volume on the various
constraints is rather simple, as shown in Figure \ref{cap:Influence-of-salt-volume}.

The size does not have a significant impact on the feedback coefficients
because the neutron spectrum changes very little with the size. The
slight evolution of the coefficient is due to the difference in neutron
escapes, which are more likely in smaller reactors.

The evolution of the breeding ratio as the salt volume increases has
two main causes: the difference in neutron escapes, and the change
in specific power, losses due to the Pa being a direct function of
the specific power. FP capture rates do not play a significant role
in this evolution. Unlike Pa, whose inventory is determined mainly
by its rapid radioactive decay, their concentration at equilibrium
depends on the reprocessing time. The longer reprocessing time exactly
compensates the effect due to the smaller specific power.

The neutron flux in the core is directly related to the specific power
and the graphite's life span varies accordingly. Similarly, the inventory
in the core depends on the volume but the effect is not directly proportional,
because neutron escapes are different.

One last important aspect is the thermal hydraulic constraint. The
thermal power is evacuated by the fuel which thus has to circulate
in the exchangers. A limit has to be set on the out of core salt volume
so as to allow reactor control: the delayed neutrons precursors migrate
away from the neutron flux along with the salt. In our studies, the
external salt volume is 1/3 of the total volume. Heat evacuation becomes
more difficult as the specific power increases. Small sized or high
power reactors are at a disadvantage in this respect.

\subsection{Influence of the Salt Composition}

\subsubsection{Elimination of the Be}

By definition, the salt plays a central role in MSRs. Serving as the
solvent for the fuel, as the moderator and as the coolant, it has
to have many characteristics specific to the neutronic as well as
the chemical, hydraulic, or thermal aspects. Like the MSRE, the MSBR
was based on a fluoride salt, because of its good neutronic properties
(capture rate and moderating capability) in a thermal spectrum. Lithium
was chosen for the same reasons and beryllium because it brought the
melting temperature down to 490~°C. The salt composition was 71.7\%~LiF
- 16\%~Be$\textrm{F}_{2}$ - 12.3\%~(HN)$\textrm{F}_{4}$ %
\footnote{~The composition that we really used in our tests with this type
of salt \cite{Alex} is: 70\% LiF - 17.5\% Be$\textrm{F}_{2}$ - 12.5\%
(HN)$\textrm{F}_{4}$.%
}.

Our first step in studying the influence of the salt composition was
to eliminate beryllium from the salt bringing it to the eutectic point
78\%~LiF - 22\%~(HN)$\textrm{F}_{4}$. All the studies discussed
up to now were done with this composition. The reasons for eliminating
beryllium are based mainly on problems with its chemistry, its toxicity,
and its availability. The proportion of heavy nuclei in the eutectic
changes drastically and its melting point increases from 490~°C to
565~°C %
\footnote{~In order to make comparisons easier and because the temperature
difference is not large, the studies with the 78\%~LiF - 22\%~(HN)$\textrm{F}_{4}$
salt were done at the same temperature as in previous studies \cite{Lecarpentier,Alex},
i.e. 630~°C.%
}. This temperature increase remains moderate and it seems manageable
with the commonly used structure materials. Because the significantly
higher proportion of Heavy Nuclei has a strong impact on the in core
inventory and on the breeding ratio, we have decided to reduce the
salt volume from 40~$\textrm{m}^{3}$ (MSBR) to 20~$\textrm{m}^{3}$
(reference configuration for these studies) so as to keep the same
amount of Heavy Nuclei in the reactor and, as a result, similar neutronic
behavior.

The elimination of the beryllium impacts all five constraints. The
presence of a ternary component in the salt seems to complicate the
reprocessing chemistry, the risk being that this element be extracted
instead of the target elements. Thus, using the LiF - (HN)$\textrm{F}_{4}$
salt could simplify fuel reprocessing and, as a result, bring it closer
to feasibility. The impact on the other constraints is shown in Table
\ref{cap:Contraintes-en-fonction-type-sel}. The spectrum is harder
with the binary salt because of the larger proportion of HN. This
translates directly into an improvement of the feedback coefficient
and a larger inventory. The change in the breeding ratio is due jointly
to the increased proportion of HN (positive action) and to the increased
specific power (negative action). The latter also implies a neutron
flux increase which, combined with the faster neutron spectrum, leads
to a significant deterioration of the graphite's life span.

\begin{table}[h]
\vspace{5mm}
\begin{center}\begin{tabular}{|c|c|c|}
\hline 
&
LiF - Be$\textrm{F}_{2}$&
LiF\tabularnewline
&
(40 $\textrm{m}^{3}$)&
(20 $\textrm{m}^{3}$)\tabularnewline
\hline
\hline 
Feedback coeff. (pcm/°C)&
-1.57&
-2.36\tabularnewline
\hline 
Breeding ratio&
1.009&
1.000\tabularnewline
\hline 
Neutron flux (x$\textrm{10}^{14}$n/$\textrm{cm}^{2}$/s)&
8.6&
13.7\tabularnewline
Graphite life span (years)&
3.3&
1.8\tabularnewline
\hline 
$^{233}\textrm{U}$ inventory (kg)&
1650&
1925\tabularnewline
\hline
\end{tabular}\end{center}

\caption{Constraints according to the type of salt used to dissolve the fuel.
The statistical error on the feedback coefficients is less than 0.05~pcm/°C
(configuration: r = 8.5~cm, variable salt volume, 630~°C, variable
(HN)$\textrm{F}_{4}$ proportion).\label{cap:Contraintes-en-fonction-type-sel}}
\end{table}

The elimination of beryllium has an additional and significant advantage,
that is not related to the constraints that we have identified in
this paper. Tritium is produced, in a system like the MSBR, by the
(n,nt) reaction on $^{7}\textrm{Li}$ and the (n,t) reaction on $^{6}\textrm{Li}$,
producing 2/3 and 1/3 respectively of the tritium \cite{Alex}. The
lithium used is 99.995 \% enriched with $^{7}\textrm{Li}$; the $^{6}\textrm{Li}$
is rapidly consumed, unless it is regenerated by an (n,$\alpha$)
reaction on $^{9}\textrm{Be}$. With the elimination of beryllium,
this reaction cannot occur and, as a result, the production of tritium
is reduced.

\subsubsection{Evaluation of the LiF - (HN)$\textrm{F}_{4}$}

\paragraph{Temperature increase :}

The proportion of Heavy Nuclei in the binary salt can be adjusted.
As it is reduced, the melting point increases, reaching 845~°C with
pure LiF. Common structure materials cannot withstand such a temperature
increase. However, new promising solutions based on carbon (carbon-carbon,
carbon fiber, carbides,...) could help solve this problem \cite{graphite}.
If this technology can be implemented, then the HN proportion parameter
can be modified.

\begin{table}[h]
\vspace{5mm}
\begin{center}\begin{tabular}{|c|c|c|}
\hline 
&
630°C&
1030°C\tabularnewline
\hline
\hline 
Feedback coeff. (pcm/°C)&
-2.36&
-1.00\tabularnewline
\hline 
Breeding ratio&
1.000&
1.026\tabularnewline
\hline 
Neutron flux (x$\textrm{10}^{14}$n/$\textrm{cm}^{2}$/s)&
13.7&
9.6\tabularnewline
Graphite life span (years)&
1.8&
1.2\tabularnewline
\hline 
$^{233}\textrm{U}$ inventory (kg)&
1925&
1630\tabularnewline
\hline
\end{tabular}\end{center}

\caption{Constraints according to the mean temperature of the fuel salt. The
statistical error on the feedback coefficients is less than 0.05 pcm/°C
(configuration: r = 8.5~cm, 20 $\textrm{m}^{3}$ salt, variable temperature,
22\% (HN)$\textrm{F}_{4}$)\label{cap:Contraintes-en-fonction-temperature}}
\end{table}

The temperature increase due to the change of salt leads us to set
the operating temperature at 1030 °C for all the configurations. We
will first study the influence of this temperature hike on the standard
configuration before studying the influence of the HN proportion in
the salt at 1030~°C.

At this temperature, the thermodynamic efficiency is assumed to increase
from 40~\% to 60~\% and this has an incidence on the thermal power
of the reactor: 1666~MWth instead of 2500~MWth are needed to produce
1000~MWe. Similarly, the salt density decreases form 4.3 to 3.89
because of the temperature related expansion effect. The impact on
the constraints of this temperature increase is detailed in Table
\ref{cap:Contraintes-en-fonction-temperature}.

The change in salt density has a direct influence on the moderation
ratio, resulting in a better thermalization of the neutron spectrum.
This induces, for a channel radius of 8.5~cm, a worsening of the
feedback coefficient (a behavior similar to that shown in Figure \ref{cap:D=E9composition-du-coefficient-fonction-rayon}).
Similarly, this slight thermalization leads to a larger $^{233}\textrm{U}$
fission cross section and, combined with the lower salt density, a
smaller necessary inventory. As for the breeding ratio, it is improved
because of the reduced specific power, which has a direct incidence
on parasitic captures (mainly those of FPs and Pa). The lower specific
power has an incidence also on the neutron flux and, thus, on the
graphite's life span. However, at such a temperature, the fluence
limit that the graphite can withstand is reduced from 2.$\textrm{10}^{22}$~n/$\textrm{cm}^{2}$
to $\textrm{10}^{22}$~n/$\textrm{cm}^{2}$ \cite{tenue_graphite}.
As a result, the graphite's life span is reduced in spite of the smaller
neutron flux.

\paragraph{Influence of the proportion of Heavy Nuclei:}

Now that the effect of the temperature change from 630~°C to 1030~°C
is known, the impact of the proportion of Heavy Nuclei can be explored.
It is useful to keep the total amount of Heavy Nuclei constant, as
we did when we changed the salt composition. As a consequence, salts
with a smaller proportion of HN will have a larger volume. This makes
these new configurations potentially interesting from the point of
view of thermal power extraction. The core reprocessing time is kept
at 6 months since, in that case, the flow of HN to be reprocessed
is the same for all the configurations. The salt of the thorium blanket
is not modified. The density and expansion coefficient of the fuel
salt are crucial parameters, they are given in Table \ref{cap:tableau Caracteristiques-en-fonction-pourcentage-NL}
\cite{Walle2}.

\begin{figure}[h]
\vspace{5mm}
\begin{center}\includegraphics[%
  clip,
  scale=0.3]{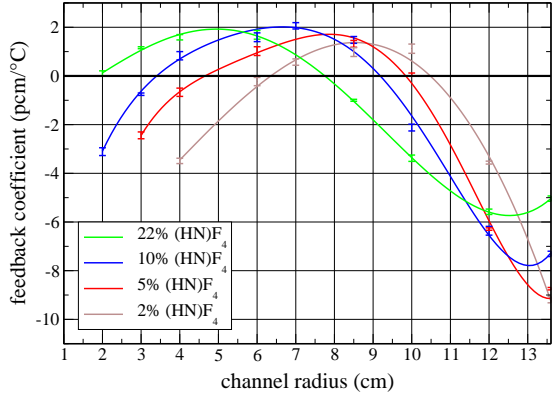}\end{center}

\caption{Feedback coefficient versus channel radius for several proportions
of (HN)$\textrm{F}_{4}$ (configuration: variable radius, variable
salt volume, 1030~°C, variable proportion of (HN)$\textrm{F}_{4}$).
\label{cap:coef_fonction_ rayon_et_proportion_NL}}
\end{figure}

\begin{figure}[h]
\vspace{5mm}
\begin{center}\includegraphics[%
  clip,
  scale=0.3]{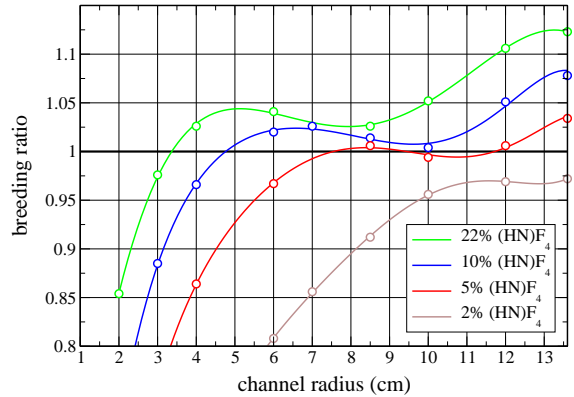}\end{center}

\caption{Breeding ratio versus channel radius for several proportions of (HN)$\textrm{F}_{4}$
(configuration: variable radius, variable salt volume, 1030~°C, variable
proportion of (HN)$\textrm{F}_{4}$).\label{cap:taux_reg_fonction_ rayon_et_proportion_NL}}
\end{figure}

Since the graphite's life span is directly related to the specific
power and since the inventory is kept constant, these constraints
are not very interesting in this part of our study. We will thus concentrate
our attention on the safety and the regeneration constraints. Rather
than presenting the impact of the proportion of HN for a reference
configuration (r = 8.5~cm, salt volume = 20~$\textrm{m}^{3}$) as
was done previously, we will look at the impact of the channel radii
for different HN proportions. This view point will allow a better
understanding of the phenomena at play. The results are shown in Figures
\ref{cap:coef_fonction_ rayon_et_proportion_NL} and \ref{cap:taux_reg_fonction_ rayon_et_proportion_NL}.

\begin{table}[h]
\vspace{5mm}
\begin{center}\begin{tabular}{|c|c|c|c|c|}
\hline 
&
22\%&
10\%&
5\%&
2\%\tabularnewline
\hline
\hline 
Salt volume ($\textrm{m}^{3}$)&
20&
36.8&
67.2&
155\tabularnewline
\hline 
Density&
3.89&
2.85&
2.33&
1.98\tabularnewline
\hline 
\begin{tabular}{c}
Expansion coeff.\tabularnewline
(x$\textrm{10}^{-4}$/°C)\tabularnewline
\end{tabular}&
10&
10&
9&
8\tabularnewline
\hline
\end{tabular}\end{center}

\caption{System properties according to the percentage of (HN)$\textrm{F}_{4}$\label{cap:tableau Caracteristiques-en-fonction-pourcentage-NL}}
\end{table}

\begin{figure}[h]
\begin{center}\includegraphics[%
  clip,
  scale=0.3]{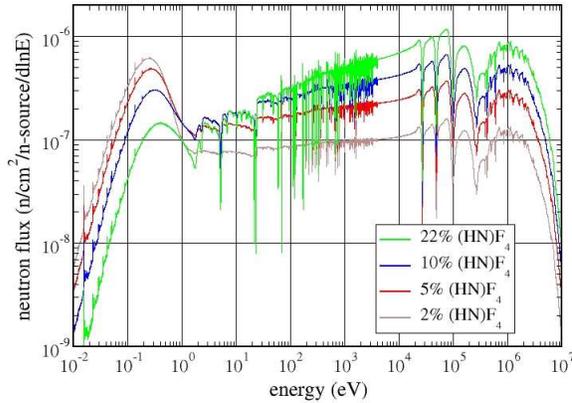}\end{center}

\caption{Neutron spectra for several proportions of (HN)$\textrm{F}_{4}$
(configuration: r = 8.5~cm, variable salt volume, 1030~°C, variable
proportion of (HN)$\textrm{F}_{4}$).\label{cap:Spectre-de-neutron-fonction_proportion}}
\end{figure}

During the reprocessing outlined in Figure \ref{cap:Reprocessing-scheme},
the thorium is extracted in order to allow the extraction of the FPs
from the salt. This step is a key point in the reprocessing and it
is made easier if the proportion of HN in the salt is small.

Moreover, when this proportion is decreased, the neutrons are scattered
for a longer time before they encounter a fissile or fertile element
(those that dominate neutron absorptions). This leads directly to
a more thermalized neutron spectrum, as shown on Figure \ref{cap:Spectre-de-neutron-fonction_proportion}.
Thus, the behavior of configurations with a small proportion of HN
is similar to that of configurations with a 22\%~(HN)$\textrm{F}_{4}$
proportion but with smaller channels.

This additional thermalization is the main cause of the evolution
of the feedback coefficients. And the visible difference for the configurations
with a single salt channel seems to be due mainly to the improvement
of the Doppler sub-coefficient. Likewise, the breeding ratio curves
are similar from one salt to the other but two effects are observed.
The first effect is related to the thermalization change and the second
to a deterioration of the breeding ratio when the HN proportion is
decreased. This is due to an increased capture rate in the light elements
of the salt on which the neutrons scatter for a longer time.

\section{General Discussion}

The various studies that have been carried out lead to a better understanding
of the way an MSR works. The search for reactor configurations, be
it a demonstrator or a power generator, requires that a certain number
of constraints be satisfied. To do so, the different studies discussed
in this paper have to be combined and the results extrapolated. Since
the parameters are not mutually independent, one has to be circumspect
in this approach.\\

Let us try to explore the possible reactor configurations. For the
sake of simplicity, the parameter concerning the salt composition
will not be considered in a first approach. The salt, then, is 78\%~LiF
- 22\%~(HN)$\textrm{F}_{4}$ and the mean temperature is 630~°C.

Since the safety aspect cannot be circumvented in the design process
of a nuclear reactor, we consider that this constraint is necessarily
satisfied. Moreover, we consider only those configurations whose total
feedback coefficient, not just the salt feedback coefficient, is negative.
Except in the case where the size of the reactor is reduced dramatically,
leading to a significantly increased neutron flux, the total feedback
coefficient is negative only for either very thermalized or fast neutron
spectra. 

The first option implies a small fissile matter inventory and a weak
neutron flux. When submitted to such a flux, the graphite undergoes
little damage and its life span is reasonably long. On the other hand,
captures in the moderator deteriorate the breeding ratio significantly.
If a reactor system does not need to regenerate its fuel, then this
very thermalized configuration may be suitable.

The faster neutron spectrum option introduces a real difficulty concerning
the graphite, whose life span is then on the order of one year. There
are several solutions to this problem.

\begin{itemize}
\item Decreasing the specific power of the reactor (by increasing its size
and/or decreasing the total power generated) leads directly to a decreased
flux intensity and, as a consequence, extends the graphite's life
span. This, however, increases in the same proportion the per GWe
fissile matter inventory, without providing a very satisfactory solution.
\item The absence of moderating graphite in the single salt channel configuration
solves this problem, the graphite in the periphery of the core being
much less irradiated. However, this option leads to the fastest neutron
spectrum and, as a result, the largest fissile matter inventory (5
to 6 metric tons of $^{233}\textrm{U}$). The specific power can be
increased, though, in order to reduce the per GWe inventory.
\item It may be possible to use a material whose structure is much less
sensitive to irradiation than graphite. Then the configuration space
for channel radii lying between 6 and 13~cm would no longer be forbidden.
\end{itemize}
Finally, the high breeding ratio obtained thanks to the fast neutron
spectrum provides more leeway for the reprocessing. In particular,
the time for full core reprocessing can be significantly extended.
The single salt channel configuration discussed earlier can even do
without any reprocessing (except for the bubbling process and uranium
retrieval in the blanket) and still regenerate its fuel during the
first 20 years of operation. The salt could conceivably be replaced,
or entirely reprocessed after this time has elapsed. If a 6 month
reprocessing time is kept, it becomes possible to do without a thorium
blanket while still regenerating the fuel. This would substantially
simplify the reactor design. Finally, we should note that, with a
fast spectrum, re-injecting the TranUranians in the core would be
interesting, both in terms of regeneration (small neutron losses)
and in terms of waste production (good incineration capability).\\

If it proves possible, increasing the temperature to above 1000~°C
has many positive repercussions on the constraints, in particular
thanks to the increased thermodynamic efficiency. However, it induces
better thermalization by the salt and this has to be taken into account.

This opens the way to salt compositions containing small amounts of
Heavy Nuclei. The way these reactors behave is practically the same
with such compositions as with 22\%~(HN)$\textrm{F}_{4}$. In particular,
the neutron spectra have to be either very thermalized or fast to
ensure negative feedback coefficients. However, this corresponds to
different reactor configurations because of the increased thermalization
due to the salt's light nuclei. The capture rate of these light elements
deteriorates the breeding ratio and leaves less leeway than with compositions
containing more HN. The fundamental importance of this parameter lies
elsewhere. Indeed, it is possible either to keep the size constant,
thus reducing the inventory but not the specific power, or to increase
the size, thus decreasing the specific power but not the inventory.
A compromise between these two extremes can be found since decreasing
the specific power facilitates the evacuation of the thermal power,
a constraint that must not be neglected.

\section*{Conclusion}

While our studies were, at first, close to the MSBR configuration,
they prompted us to diversify our investigations. We analyzed the
impact on the behavior of the core of such parameters as the reprocessing,
the moderation ratio, the core size, and the proportion of heavy nuclei
in the salt.

Our results confirm that there is a problem with the feedback coefficients
in the MSBR. In a thermal spectrum, it would be possible to reach
an acceptable concept only after in depth investigations taking into
account the effect of the salt (negative feedback coefficient) and
of the graphite (which makes the global feedback coefficient positive)
separately. For very thermalized spectra, the global coefficients
are negative thanks to the large neutron losses in the moderator but
this leads also to a very poor breeding ratio. Epithermal or fast
neutron spectra thus seem more favorable since they combine good feedback
coefficients with satisfactory breeding ratio. However they lead to
severe problems with the graphite's ability to withstand the irradiation.
As a result, the solution that removes the moderating block seems
especially attractive.

Our studies have uncovered a wider range of possibilities than anticipated.
Thus, many options remain to be explored. In particular, the evaluation
of new materials, be it to obtain a moderator that has better irradiation
resistance properties or to allow high temperature operation is crucial
for an even more interesting development of the concept. How to extract
the thermal power from the core is another issue of major interest
since it impacts the behavior of the core through the specific power
aspect. In order to ease heat recovery, the salt composition can also
be modified so as to dissolve the fissile matter in a larger salt
volume. In general, it is possible to change the type of salt, the
MSR concept being adaptable to such a change.

As many parameters remain to be studied, other acceptable solutions
could be found. In particular, parameters such as the type of salt,
the moderating material, the size of the lattice hexagons, the definition
of several different areas in the core, could be studied more specifically.
In view of the results already obtained, it is clear that many configurations
remain to be explored, requiring research on the salt and the materials
as well as on the neutronics and the geometry.

\section*{Acknowledgements}

The paper took benefit from the works made with M.Allibert and discussions
during the european contract MOST (MOlten SalT : FIKW-CT-2001-20183).
We would like to thank Elisabeth Huffer for her translation (from
French to English) of this paper.

\end{document}